\documentstyle[epsfig]{jfm}

\catcode`\œ=\active \gdefœ{\setbox0=\hbox{0}\hbox to\wd0{}}
\ifCUPmtlplainloaded
\else
\fi
\ifCUPmtlplainloaded

  \renewcommand{\simeq}{\approx}
\fi
\ifCUPmtlplainloaded
  \font\bit = mtmib10 at 10.5pt \skewchar\bit ='177  
\else
  \font\bit = cmmib10 \skewchar\bit ='177  
\fi
\ifCUPmtlplainloaded
\else
  \font\tenbmi=cmmib10 at 10pt  \skewchar\tenbmi ='177
  \font\sevenbmi=cmmib10 at 7pt \skewchar\sevenbmi ='177
  \font\fivebmi=cmmib10 at 5pt  \skewchar\fivebmi ='177

  \newfam\bmifam
  \textfont\bmifam=\tenbmi
  \scriptfont\bmifam=\sevenbmi
  \scriptscriptfont\bmifam=\fivebmi
  
\fi
\newsavebox{\thalfbox}
\sbox{\thalfbox}{$\textstyle\frac{1}{2}$}

\newsavebox{\shalfbox}
\sbox{\shalfbox}{$\scriptstyle\frac{1}{2}$}

\newsavebox{\squartbox}
\sbox{\squartbox}{$\frac{1}{4}$} 

\newsavebox{\etbox}
\sbox{\etbox}{\boldmath$\eta$}

\newsavebox{\astrutbox}
\sbox{\astrutbox}{\rule[-5pt]{0pt}{20pt}}

\ifnfsstwo

\fi
\ifnfssone
  \newmathalphabet{\mathit}
    \addtoversion{normal}{\mathit}{cmr}{m}{it}
    \addtoversion{bold}{\mathit}{cmr}{bx}{it}

\fi
\ifoldfss

\fi

\mathchardef\varLambda="0103

\ifCUPmtlplainloaded
\else
\fi
\ifCUPmtlplainloaded
  \let\bcdot=\undefined
  \NewSymbolFont{bldsym}{mtbsy10}{'60}
  \NewMathSymbol{\bcdot}{2}{bldsym}{01}
\else
  \font\tenbms=cmbsy10          \skewchar\tenbms ='60
  \font\sevenbms=cmbsy10 at 7pt \skewchar\sevenbms ='60
  \font\fivebms=cmbsy10 at 5pt  \skewchar\fivebms ='60

  \newfam\bmsfam
  \textfont\bmsfam=\tenbms
  \scriptfont\bmsfam=\sevenbms
  \scriptscriptfont\bmsfam=\fivebms

  \edef\bsy{\hexnumber\bmsfam}
  \mathchardef\bnabla="0\bsy72
  \mathchardef\bcdotsymbol="0\bsy01
  \def\bcdot{\,\bcdotsymbol\,}
\fi
\title[]{Scale by scale budget and similarity laws for shear turbulence}

\author[C\@. M\@. Casciola, P\@. Gualtieri,  R\@. Benzi and R\@. Piva]
{
C.\ls M.\ns C\ls A\ls S\ls C\ls I\ls O\ls L\ls A,$^1$ \ns 
P.\ns G\ls U\ls A\ls L\ls T\ls I\ls E\ls R\ls I,$^1$ \ns
R.\ns B\ls E\ls N\ls Z\ls I,$^2$  \\ \ns 
\and R.\ns P\ls I\ls V\ls A$^1$}

\affiliation{$^1$Dipartimento di Meccanica e Aeronautica, \UNIV di Roma {\em{La Sapienza}},
Via Eudossiana 18, 00184 Roma, {\it{Italy}}\\[\affilskip]
$^2$Dipartimento di Fisica e INFM, \UNIV di Roma {\em{Tor Vergata}},
Via della Ricerca scientifica 1, 00133 Roma, {\it{Italy}}}

\pubyear{2002}
\volume{1}
\pagerange{1--10}
\date{$\tilde{\quad}$}
\newcommand {\UNIV}   {Universit\`a }   
\newcommand {\ds}   {\displaystyle}
\newcommand {\bce}  {\begin{center}}
\newcommand {\ece}  {\end{center}}
\newcommand {\be}   {\begin{equation}}
\newcommand {\ba}   {\begin{array}}
\newcommand {\bea}  {\begin{eqnarray}}
\newcommand {\bfi}  {\begin{figure}}
\newcommand {\ee}   {\end{equation}} 
\newcommand {\ea}   {\end{array}}
\newcommand {\eea}  {\end{eqnarray}}
\newcommand {\efi}  {\end{figure}}
\newcommand {\noi}  {\noindent}

\def\Rset {{\rm I \kern-.2em R}} 
\def\mathbbH {{\rm I \kern-.2em H}} 
\def\mathbbC {{\rm I \kern-.6em C}} 

\begin{document}
\maketitle
\begin{abstract}
Turbulent shear flows, such as those occurring in the wall region of turbulent 
boundary layers, manifest a substantial increase of intermittency with respect to 
isotropic conditions.
This suggests a close link between anisotropy
and intermittency.  However, a rigorous statistical description of anisotropic 
flows is, in most cases, hampered by the inhomogeneity of the field.
This difficulty is absent for the homogeneous shear flow.
For this flow the scale by scale budget is discussed here by using the appropriate  
form of the Karman-Howarth equation, to determine the range of 
scales where the shear is dominant. 
The issuing generalization of the {\sl four-fifths} law 
is then used as the guideline to extend to shear dominated flows the
Kolmogorov-Obhukhov theory of intermittency.
The procedure leads naturally to the formulation of generalized
structure functions and the description of intermittency thus obtained
reduces to the K62 theory for vanishing shear.
Also here the intermittency corrections to the scaling exponents are carried
by the moments of the coarse grained energy dissipation field.
Numerical experiments give indications that 
the dissipation field is statistically unaffected by the shear,
thereby supporting the conjecture that the intermittency corrections are 
universal.
This observation together with the present reformulation of the theory 
gives reason for the increased intermittency 
observed in the classical longitudinal velocity increments.
\end{abstract}

\section{Introduction}
At large Reynolds number, turbulent flows exhibit strong intermittent effects which
can be described in a number of ways, such as intense coherent structures, 
anomalous scaling in the inertial range, exponential tails in the probability density 
functions (see e.g. \cite{frisch}).
In the case of homogeneous and isotropic turbulence, more than twenty years of intense 
scientific research provide us with strong indications that intermittency shows universal
properties (\cite{sreeni}).

As soon as we consider non homogeneous non isotropic turbulence, 
a well established conclusion concerning universal properties 
of intermittency is still missing, although some preliminary results have been proposed. 
See for instance the papers of \cite{toschi_1}, \cite{danaila_jfm},
\cite{bife}.
A particular case 
of non isotropic turbulence is provided by homogeneous shear flows which have been
the subject of a number of numerical (see \cite{pumir} and references therein)
and experimental investigations (\cite{shen}). 
Homogeneous shear flows are presumably the simplest non trivial example of non isotropic
turbulence where issues such as the generation and the dynamics of coherent 
vortical structures, the effect of the large scale non isotropic forcing on the small scale 
intermittent fluctuations of velocity and the universal (if any) scaling properties of 
velocity increments in the inertial range can be addressed.

In the present paper we investigate intermittency effects in homogeneous shear 
flows by using a long time highly resolved Direct Numerical Simulation (DNS), see \cite{PF_paolo} 
for details. Our basic idea
is to study the fluctuations of the velocity field by using a generalization of the 
Kolmogorov {\em{four-fifths}} equation. Although for the homogeneous shear flow 
no use of the isotropic constraints can be made, we are still able to write the 
appropriate form of this equation following the original idea due to \cite{hinze}. 
As an advantage, we can immediately isolate the contributions due to the mean 
shear to identify the range of scales where they become the leading terms. 
Next, by generalizing the concept of structure functions, we are able to study 
systematically the scaling properties of shear turbulence and its universality
with respect to homogeneous and isotropic conditions.

The Kolmogorov equation for homogeneous isotropic turbulence follows from 
the Karman-Howarth equation
\be
\label{4_5_intro}
    <\delta V_\parallel^3> = -\frac{4}{5} \bar{\epsilon} r + 6 \nu \frac{d}{dr}
    <\delta V_\parallel^2>
\ee
where 
$
\delta V_\parallel = [\vec u(\vec x + \vec r) - \vec u(\vec x)
]\cdot {\vec r}/ {r}
$
are the longitudinal velocity increments. 
In the inertial range, the Karman-Howarth equation reduces to the well known {\em{four-fifths}}
law relating the third order structure function to the mean rate of energy dissipation
$\bar{\epsilon}$ and to the separation $r$. In absence of intermittency effects the
{\em{four-fifths}} law suggests that, beyond the separation $r$, any inertial range scalings
should involve only the mean rate of energy dissipation $\bar{\epsilon}$,
or, more properly, the average energy flux through the inertial range.
In other words, scalings in terms of separation for structure 
functions of order $p$ should necessarily be given by
\be
\label{dim_scaling}
    <\delta V_\parallel^p> \propto \bar{\epsilon}^{p/3} r^{p/3} \, .
\ee
It follows that the dimensionless ratios of structure functions must go to a
constant as the separation is reduced within  the inertial range 
\be
\label{dim_ratio}
    \lim_{r \rightarrow 0} {<\delta V_\parallel^p>} / {<\delta V_\parallel^2>^{p/2}}
    = const \, ,
\ee
\noi i.e. no intermittency is present.
This is essentially the K41 theory of homogeneous isotropic turbulence (see Frisch).
Preliminarily we may observe that eq.~(\ref{dim_ratio}) implies 
$<\delta V_\parallel^p> \propto <\delta V_\parallel^q>^{p/q}$.

When intermittency is accounted for, eq. (\ref{dim_ratio}) is not valid.
As a matter of fact scaling laws are nonetheless observed in the inertial range, 
with exponents substantially different from the dimensional prediction for large
$p$ (\cite{benzi_hi}).
The observed behavior can be recast in the form
\be
\label{scaling_exp}
    <\delta V_\parallel^p> \propto \bar{\epsilon}^{p/3} r^{p/3} 
\left(\frac{r}{L_0}\right)^{\tau(p/3)},
\ee
where $L_0$ is the integral scale of turbulence and $\tau(p/3)$ are the intermittency 
corrections to the dimensional scaling exponents. 

In the context of the Kolmogorov Refined Similarity Hypothesis (RKSH)
(\cite{kolm_62}) the intermittency corrections are associated to the statistical properties
of the coarse grained energy dissipation,
\be
\label{eps_r_def}
    \epsilon_r=\frac{1}{B_r} \int_{B_r} \epsilon_{loc} \quad dV_r \, ,
\ee
with $\epsilon_{loc}$ the local rate of energy dissipation, by expressing the structure
functions as
\be
\label{k_62}
    <\delta V_\parallel^p> \propto <\epsilon_r^{p/3}> r^{p/3}.
\ee
In words,
the intermittency of the velocity field is originated by an intermittent energy cascade 
which ends up in an intermittent dissipation field. 
The link between the two fields is provided by eq. (\ref{k_62}) where
\be
\label{eps_r_p}
    <\epsilon_r^{p/3}> \propto r^{\tau(p/3)}.
\ee
Intermittency effects on velocity structure functions in the inertial
range are documented beyond any doubt (see e.g. \cite{zhou}).
More controversial is the assessment of eq. (\ref{k_62}).
However the evidence is all in favor of this equation, which, 
though not derived from first principles,  seems  to describe correctly the physics of 
small scale intermittency (\cite{wang}). 
In this context, the {\em four-fifths} law is an exact constraint to be fulfilled by any
acceptable theory of turbulence.
Besides this, the Kolmogorov and the Karman-Howarth equations may also suggest the 
basic ingredients of a reasonable turbulence theory.

We can now re-express more technically the aim of the present paper. Namely, 
starting from the extension of the Karman-Howarth equation to the homogeneous shear
flow (see also \cite{Oberlack} for a related derivation), we are interested in generalizing 
the RKSH to anisotropic turbulence. 
As we shall see, the proposed similarity law includes the classical Kolmogorov-Obhukhov  
theory of intermittency as a special case. 
Further, we intend to check the proposed form of similarity by using DNS results of a 
homogeneous shear flow. 
This may make things appear a little less straightforward, since, given the limitations 
of DNS, scaling laws cannot be detected directly in terms of separation. 
To overcome the difficulty we analyze our data {\sl via} Extended Self Similarity 
(\cite{benzi_ess}), to show that, indeed, the generalized RKSH holds for the homogeneous shear flow.
Actually ESS scalings, 
$<\delta V_\parallel^p> \propto <\delta V_\parallel^q>^{\zeta(p|q)}$ with 
$\zeta(p|q) \ne p/q$, are observed 
also at moderate Reynolds number both for isotropic and for shear turbulence 
(see \cite{antonia}, \cite{ruiz} and \cite{toschi_1}).  
In particular, for homogeneous isotropic turbulence and within the ESS rationale 
one can rewrite eq.~(\ref{k_62}) as
\be
\label{k_62_ESS}
    <\delta V_\parallel^p> \propto {<\epsilon_r^{p/3}>}/{\bar{\epsilon}^{p/3}}
    <\delta V_\parallel^3>^{p/3}\, ,
\ee

\noi to check its validity against moderate Reynolds number data
(\cite{benzi_rksh}).

The generalization of the Kolmogorov-Obhukhov
theory proposed here is well supported by the data and rationalizes
a number of recent results concerning intermittency in 
shear dominated turbulence.
Actually a number of papers have recently appeared on this subject.
Strictly related to the present results is the paper by
\cite{PF_mamma} where a new form of similarity law has been proposed
to explain the increased intermittency of the near wall region of wall bounded flows
noticed by a number of authors (see \cite{ruiz}, \cite{toschi_1}).
That form of similarity law has been checked by using experimental 
(Jacob et {\em al.}) and DNS data
(Gualtieri et {\em al.}), to finally address the possible cohexistence of 
two different intermittent regimes (Casciola et {\em al.}).
In the present paper we discuss a unifying formulation
showing that the classical RKSH and the form proposed by \cite{PF_mamma} are obtained 
as limiting cases of a more general similarity law.

Despite the fact that the Karman-Howarth equation for
homogeneous isotropic turbulence is conveniently expressed in terms of third and second order
longitudinal structure functions, more generally it involves integrals on the
sphere of radius $r$ such as
\be
\label{dv_sphere}
    \oint_{B_r} <\delta V^2 \delta V_\parallel> \quad dS_r
\ee
etc. 
The primary form of Karman-Howarth equation is recast in the more familiar form 
(\ref{4_5_intro}) by using isotropy.
Clearly, the last step can not be applied to anisotropic turbulence. 
Consistently the generalized Karman-Howarth equation gives a relation between integrals
on the sphere, implying that the generalized RKSH will also maintain the same flavor.


\section{Karman-Howarth equation}
In terms of fluctuating components, $\vec u$,  
a homogeneous field $\vec v$ may be written as
\be
\label{hom_flow}
    v_i = u_i + \frac{\partial U_i}{\partial x_k} x_k \, .
\ee
Here the velocity field obeys the incompressible Navier-Stokes equations and
the mean component is solenoidal, $\partial U_i / \partial x_i = 0$.
Standard notations are used for quantities evaluated at $\vec x$ and a 
prime denotes those at $\vec y = \vec x + \vec r$.
After multiplying the equation for $u_i$ 
by $u^\prime_j$ and that for $u^\prime_j$ by $u_i$, adding the results and averaging 
we obtain the equation for the two points correlation tensor
$R_{i,j}(\vec r,t)=<u_i u^\prime_j>$. The equation for the trace $R_{i,i}$ reads
\begin{eqnarray}
\label{R_ii}
\frac{\partial R_{i,i}}{\partial t} + \frac{\partial}{\partial r_k}
[<u_i u^\prime_i u^\prime_k> - <u_i u^\prime_i u_k>] + 
(U^\prime_k - U_k) \frac{\partial R_{i,i}}{\partial r_k}
\nonumber & = & \\
-\frac{\partial U_i}{\partial x_k} (R_{i,k}+R_{k,i}) +
2 \nu \frac{\partial^2 R_{i,i}}{\partial r_k \partial r_k },
\end{eqnarray}
where, by homogeneity, 
$\partial / \partial x_k <\bullet> = - \partial / \partial r_k <\bullet>$ and
$\partial / \partial y_k <\bullet> =  \partial / \partial r_k <\bullet>$, and the
pressure term drops due to incompressibility. Equation~(\ref{R_ii})
can be re-expressed in terms of structure functions, by considering the velocity 
increments
\be
\label{dv_def}
\left\{
\ba{l}
    \ds  \delta V_i(\vec x,\vec r)= u_i(\vec x + \vec r) - u_i(\vec x) \\ \\
    \ds  \delta V_\parallel(\vec x,\vec r)= \delta V_i
          \frac{r_i}{r},
\ea
\right.
\ee
$\delta V^2=\delta V_i \delta V_i$ and $\delta U_k = U^\prime_k - U_k$.
Actually, since 
$ R_{i,i}=<u_i u_i> - \frac{1}{2} <\delta V^2> $
and 
$ R_{i,k} + R_{k,i}=<u_i u_i> - <\delta V_i \delta V_k>  $,
and considering the standard manipulations 
\be
\label{dv2_dv_k}
<\delta V^2 \delta V_k> = -2 [<u_i u^\prime_i u^\prime_k>-<u_i u^\prime_i u_k>] +
<u_i u_i u^\prime_k>-<u^\prime_i u^\prime_i u_k> \, ,
\ee
the Karman-Howarth equation for the velocity increments follows as 
\be
\label{k_h_diff}
\frac{\partial}{\partial r_k}<\delta V^2 \delta V_k + \delta V^2 \delta U_k> 
+ 2 \frac{\partial U_i}{\partial x_k} <\delta V_i \delta V_k> =
4 \frac{\partial U_i}{\partial x_k} <u_i u_k> + 
2 \nu \frac{\partial^2}{\partial r_k \partial r_k} <\delta V^2>,
\ee
\noi where use has been made of incompressibility.
Here the first term in the right hand side is identified 
with the average dissipation rate, given the 
balance between production of turbulent kinetic energy and viscous dissipation,
\be
\label{energy_eq}
\frac{\partial U_i}{\partial x_k}<u_i u_k>=
-\nu <\frac{\partial u_i}{\partial x_k} \frac{\partial u_i}{\partial x_k}>=
-\bar{\epsilon} \, .
\ee
All averaged terms in eq.~(\ref{k_h_diff}) are 
only function of separation and integration 
over the sphere of radius $r$ yields
\begin{eqnarray}
\label{k_h_sphere}
\oint_{\partial B_r} <\delta V^2 \delta V_\parallel> + <\delta V^2 \delta U_\parallel > \quad d S_r +
\int_{B_r} 2 S \frac{\partial U_i}{\partial x_k} <\delta V_i \delta V_k> \quad d V_r 
\nonumber & &  \\
 = -4 \bar{\epsilon} r + \frac{d}{dr}  \oint_{\partial B_r} 2 \nu <\delta V^2 > \quad d S_r \, .
\end{eqnarray}
When $\vec U = (Sy, \, 0, \, 0)$, by denoting by
$n_1$ and $n_2$ the components of the normal 
in the directions of mean flow and mean gradient, respectively, we finally arrive at
\begin{eqnarray}
\label{k_h}
\frac{1}{4 \pi r^2} \oint_{\partial B_r} <\delta V^2 \delta V_\parallel> 
+n_1 n_2 S r <\delta V^2 > \quad d S_r 
\nonumber & + &  \\
 \frac{1}{4 \pi r^2} \int_{B_r} 2 S <\delta u \delta v >\quad d V_r  =
-\frac{4}{3} \bar{\epsilon} r + \frac{1}{4 \pi r^2}
\frac{d}{dr} \oint_{\partial B_r}  2 \nu <\delta V^2> \quad d S_r \, .
\end{eqnarray}

\section{Scale by scale budget}
In homogeneous isotropic turbulence the Karman-Howarth equation describes 
the process of energy cascade towards small scales and establishes 
the balance between energy flux and dissipation according to the 
{\em{four-fifth}} law (\ref{4_5_intro}).
The energy, injected at the integral scale $L_0$, is simply transfered across the inertial 
range to be dissipated at scales of the order of the Kolmogorov length $\eta$. 
Equation~(\ref{k_h}) extends this result to anisotropic flows where the cascade process is strongly
modified by the continuous energy injection associated to the mean velocity gradient.

Following the derivation of the previous section, the non linear terms generate two
distinct contributions. In particular the correlation
\be
\label{dv3_tr}
<\delta V_{tr}^3> =\frac{1}{4 \pi r^2} \oint_{\partial B_r}<\delta V^2 \delta V_\parallel> +
n_1 n_2 S r <\delta V^2> \quad d S_r
\ee
where the subscript {\em{tr}} stays for transfer, is 
the proper generalization of the third order structure function 
$<\delta V_\parallel^3>$ to non isotropic flows. Its meaning 
is rather clear: the flux across the surface $\partial B_r$ arise due 
to transport of turbulent kinetic energy operated by both
the mean flow $S r$ and by the turbulent fluctuations $\delta V_\parallel$. 
It arises from the terms in divergence form in eq.~(\ref{k_h_diff}), 
as appropriate for a flux. 
The second term in the left hand side of eq.~(\ref{k_h}), namely
\be
\label{dv3_pr}
<\delta V_{pr}^3> =\frac{1}{4 \pi r^2} \int_{B_r} 2 S <\delta u \delta v> \quad d V_r,
\ee
where the subscript {\em{pr}} stays for production, describes 
the rate of turbulent kinetic energy production up to the scale $r$ and keeps
its form as a volume integral.

All the terms appearing in eq. (\ref{k_h}) are plotted in figure \ref{eq_bal_1}
\bfi[t!]
   \centerline{
   \epsfig{figure=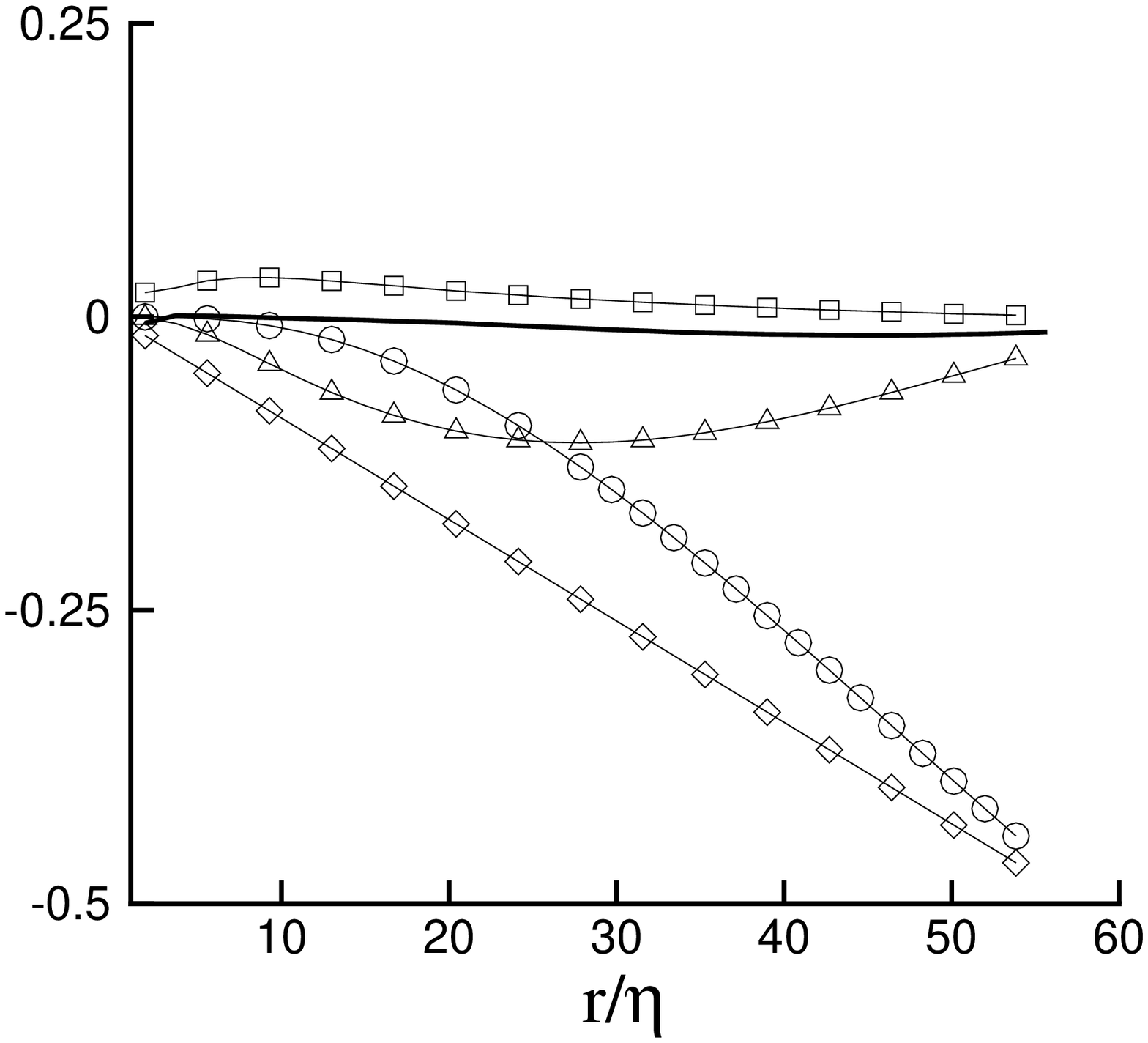,width=7.cm}
   \epsfig{figure=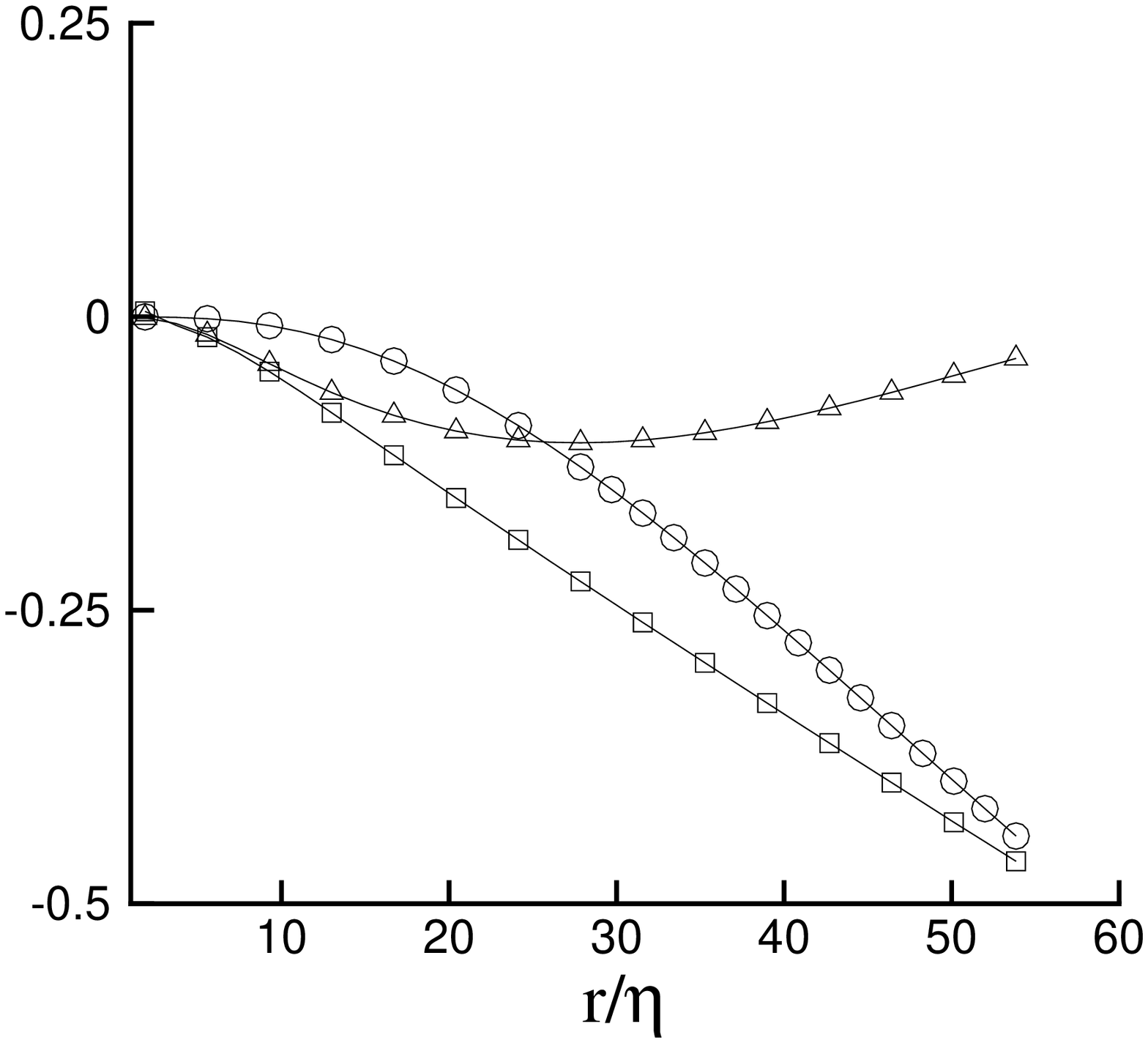,width=7.cm}}
   \caption{ Left: The different terms of eq.~(\ref{k_h}) are plotted against $r/\eta$.
    $<\delta V_{tr}^3>$, triangles, 
    $<\delta V_{pr}^3>$, circles,  
    the viscous correction, squares, and 
    $-4/3 \bar{\epsilon}r $  diamonds. The four terms sum to zero, solid line.
    Right: The squares denote $-4/3 \bar{\epsilon}r + 
    d/dr \oint_{\partial B_r} 2 \nu <\delta V^2> \, d S_r$, the other symbols are 
    defined as for the left part of the figure.
\label{eq_bal_1}}
\efi
and clearly sum up to zero for all separations $r$.
In the right part of the figure, transfer $<\delta V_{tr}^3>$ and production 
$<\delta V_{pr}^3>$ are compared to 
$ - 4 / 3 \bar{\epsilon} r$ deprived of the finite Reynolds number contributions. 
At the large scales $\bar{\epsilon}$ is balanced by 
production.
At small scales, instead, $\bar{\epsilon}$ is balanced by energy transfer.
Actually, moving towards the scales of the classical inertial 
sub-range from above,
the production having place in the larger scales piles up and originates the flux of 
energy which, transfered down the cascade, is eventually dissipated.
In fact the production and the transfer terms achieve the same order of magnitude at the
characteristic length scale where $<\delta V_{tr}^3> \simeq <\delta V_{pr}^3>$. 
The cross over scale may be estimated on dimensional grounds by assuming 
$<\delta V_{tr}^3> \simeq \bar{\epsilon} r$ and
$<\delta V_{pr}^3> \simeq S r \delta u \delta v$ with 
$\delta u \simeq \delta v = {\bar{\epsilon}}^{1/3} r^{1/3}$, to yield
the classical expression of the shear scale
$
L_s=\sqrt{{\bar{\epsilon}}/{S^3}} 
$.
This is actually well supported by the data shown in figure \ref{eq_bal_1} where the 
balance is achieved at $r_b /\eta \simeq 25$, roughly corresponding to the dimensional 
prediction of $L_s /\eta  \simeq 20$.

%

\section{Similarity laws}
 
The generalized Karman-Howarth equation~(\ref{k_h}) described in the previous sections,
in the limit of large Reynolds numbers, reduces to

\be
\label{dv_gen_r}
   < \widetilde{\delta V^3} > \propto \bar{\epsilon} r.
\ee
\noi where 
\be
\label{dv_gen}
    \widetilde{\delta V^3} = \delta V_{tr}^3 + \delta V_{pr}^3 \, .
\ee

\noi This equation is the analogous for the shear flow of the {\em four-fifths} law
of homogeneous isotropic turbulence, to which it nicely reduces for zero shear.
On purely dimensional grounds, it should generalize to higher order moments as
$
\label{dv_p_41}
   < \widetilde{\delta V^p} > \propto \bar{\epsilon}^{p/3} r^{p/3}
$.
To incorporate intermittency corrections we follow the line of reasoning that led to the 
similarity law (\ref{k_62}) by considering 
the coarse-grained dissipation field, 
eq.~(\ref{eps_r_def}), as the carrier of the intermittency corrections to the dimensional 
scalings of structure functions. In our case this procedure leads to the prediction 
\be
\label{dv_gen_k62}
    <\widetilde{\delta V^p}> \propto <\epsilon_r^{p/3}> r^{p/3}.
\ee

Unfortunately we cannot validate equation~(\ref{dv_gen_k62}) directly,
since, given the limitations in the Reynolds number of our DNS, no power-law behavior can be
observed when using the separation $r$ as scaling variable.
However, the concept of Extended Self-Similarity can help overcoming the difficulty
by recasting our ansatz in the form
\be
\label{dv_gen_k62_ext}
    <\widetilde{\delta V^p}> \propto {<\epsilon_r^{p/3}>}/{\bar{\epsilon}^{p/3}}
<\widetilde{\delta V^3}>^{p/3}.
\ee
Whenever equation~(\ref{dv_gen_k62}) holds, equation~({\ref{dv_gen_k62_ext}}) follows
by using (\ref{dv_gen_r}) to eliminate $r$ in favor of $ <\widetilde{\delta V^3}>$,
though ESS should work under more general
conditions than (\ref{dv_gen_k62}).

\bfi[t!]
   \centerline{
   \epsfig{figure=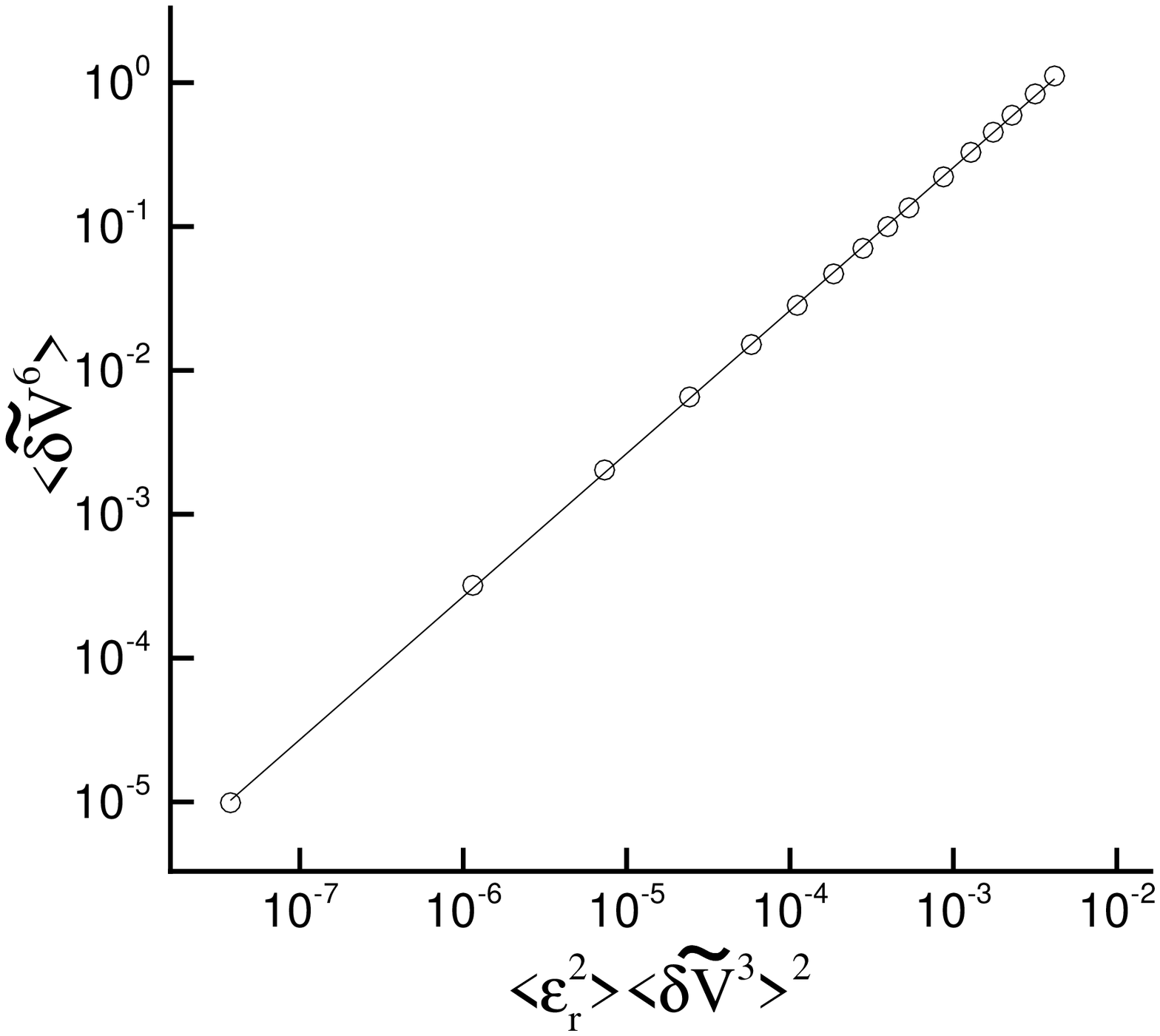,width=7.cm}
   \epsfig{figure=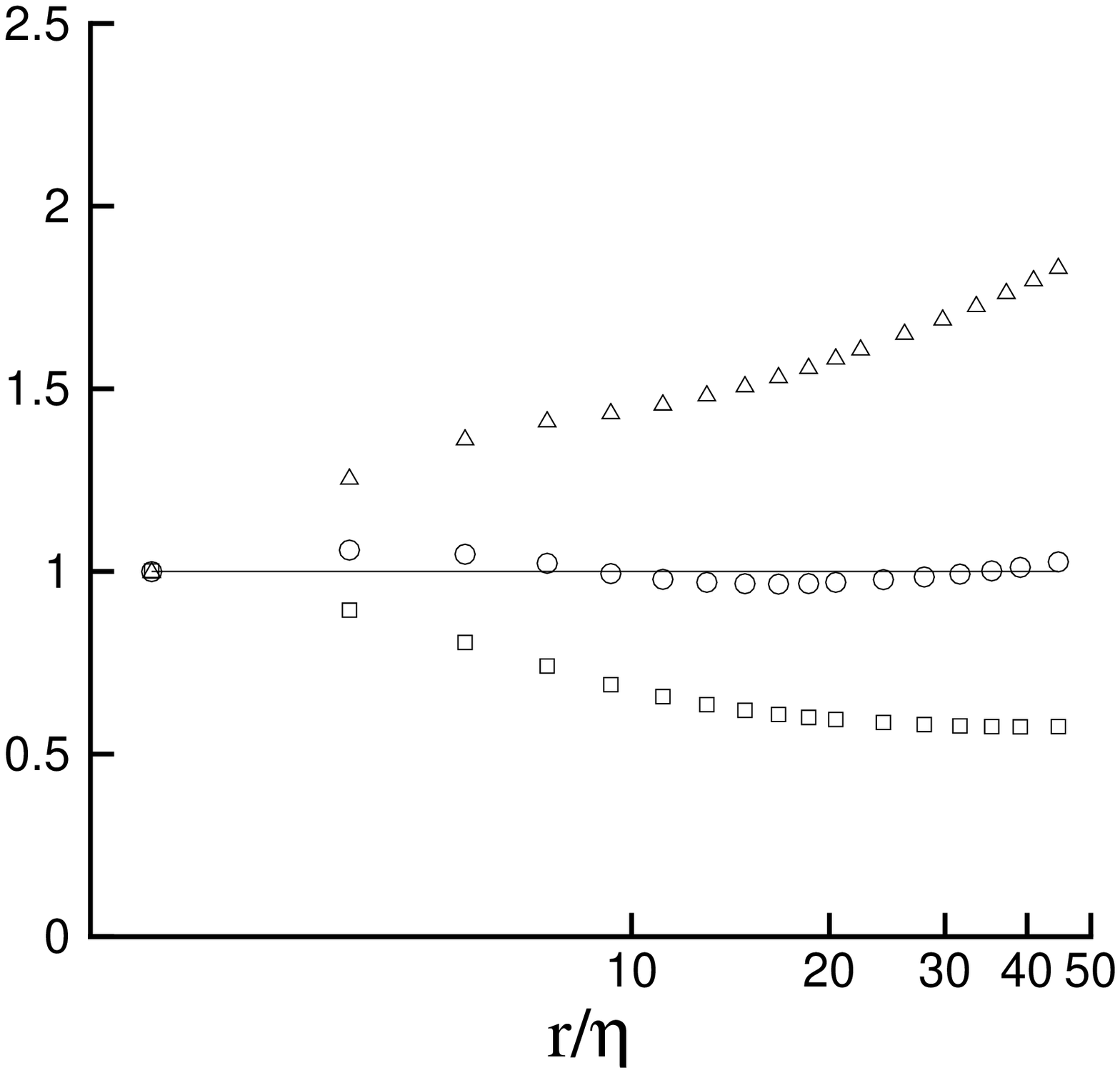,width=7.cm}}
   \caption{ Left: $<\widetilde{\delta V^6}> vs. <\epsilon_r^2> <\widetilde{\delta V^3}>^2$
            (open circles). Data are fitted in the whole range of scales
            by a power law with slope $s=1$ (solid line). The shear scale 
            approximately corresponds to the sixth point from the left.
            Right: The ratio
             $<\widetilde{\delta V^6}>/(<\epsilon_r^2> <\widetilde{\delta V^3}>^2)^\alpha$
             vs. separation $r/\eta$ for 
             $\alpha=1.05$ (triangles), $\alpha=1.00$ (circles), $\alpha=0.95$ (squares).
\label{mix}}
\efi
As shown in figure \ref{mix} for $p=6$, the data we have available well support 
equation~(\ref{dv_gen_k62_ext}) which apparently holds uniformly through the scales.
This is a peculiar feature of the present description, where 
all aspects of shear turbulence are retained, both the production term associated 
to $\delta V_{pr}^3$ and the transfer term associated to $\delta V_{tr}^3$. 
The two contributions are evaluated exactly avoiding  any 
approximation.

As a further check, we have also reported in figure \ref{mix} the ratio 
$<\widetilde{\delta V^6}>/(<\epsilon_r^2> <\widetilde{\delta V^3}>^2)^\alpha$ for 
different $\alpha$'s. 
The value  $\alpha=1$, giving  eq.~(\ref{dv_gen_k62_ext}), achieves
the optimal compensation around a constant for the whole range of scales. 
Quantitatively, the compensated data for eq.~(\ref{dv_gen_k62_ext})
fluctuate about $4.5\%$ giving clear evidence in favor of the proposed similarity law
above and below the shear scale $L_s$. 

According to eq.~(\ref{dv_gen_k62_ext}), the random function $\widetilde{\delta V^3}$ is 
statistically equivalent to the coarse grained energy dissipation $\epsilon_r$.
It is a fact that, for the present data where the shear affects 
the turbulent fluctuations enough to break the classical
refined similarity, the statistics of the dissipation field shows no appreciable 
difference with homogeneous isotropic turbulence. 
In figure \ref{eps_r_cfr} we
compare the second order moment of $\epsilon_r$ for the homogeneous shear flow
(solid line) and for homogeneous isotropic turbulence (triangles). 
The correspondence between the two statistics is confirmed 
by analyzing higher order moments, such as the third order one
plotted on the right of the same figure.
\bfi[t!]
   \centerline{
   \epsfig{figure=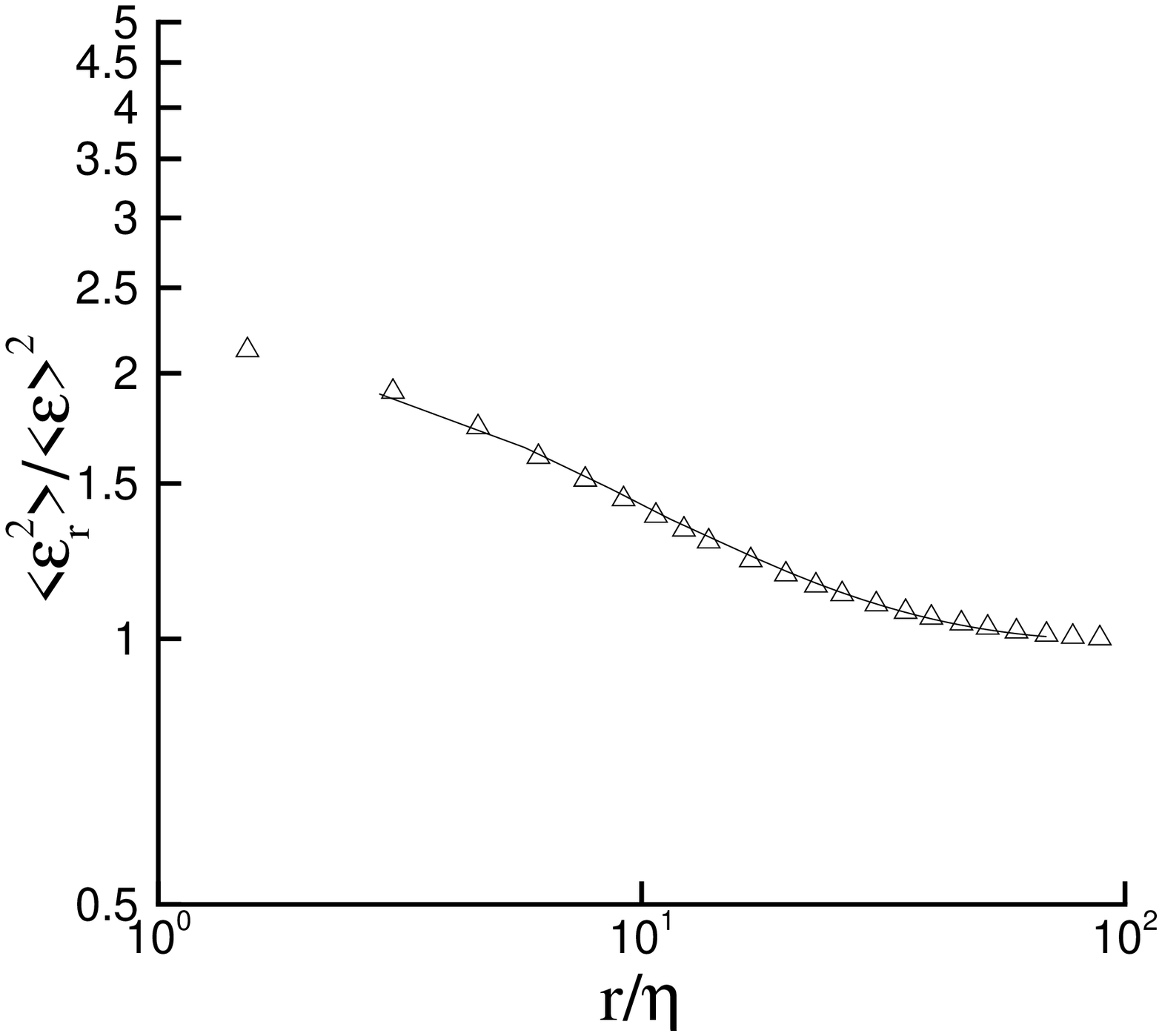,width=7.cm}
   \epsfig{figure=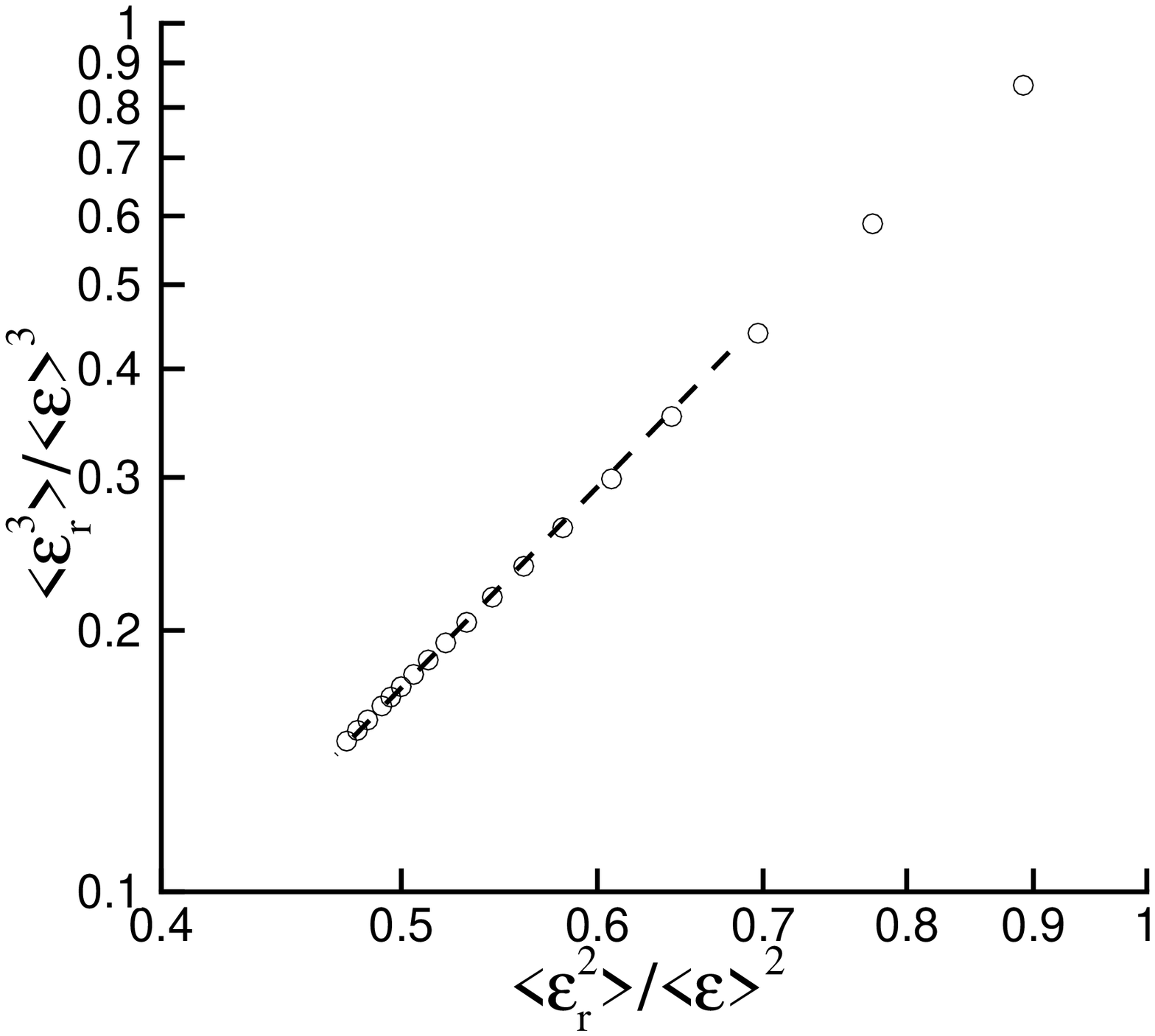,width=7.cm}}
   \caption{ Left: $<\epsilon_r^2>$ vs. $r/\eta$ in homogeneous isotropic turbulence (triangles)
             and in the homogeneous shear flow (solid line). Right:
             $<\epsilon_r^3> vs. <\epsilon_r^2>$ in homogeneous isotropic turbulence 
             (dashed  line, slope $s=2.92$)
             and in the homogeneous shear flow (open circles). 
\label{eps_r_cfr}}
\efi

The above observation and eq.~(\ref{dv_gen_k62_ext}) allow for the conclusion that the 
generalized structure functions $< \widetilde{\delta V}^p> $ are unaffected by the shear,
see the left part of figure~\ref{old_new}.
Actually, the relative scaling exponents of 
$<\widetilde{\delta V^p}> vs. <\widetilde{\delta V^3}>$ are, within $1\%$ accuracy, 
the same found in isotropic conditions. 
Under this respect, they extend to shear dominated flows the 
universal behavior well documented in homogeneous isotropic turbulence.

As an additional comment, we like to compare the present findings with related work
concerning the similarity law proposed by \cite{PF_mamma} for shear
dominated flows and the double intermittent behavior in shear turbulence
discussed by \cite{PRE}.
For $r<L_s$, the budget provided by the Karman-Howarth 
equation suggests that $\delta V_{tr}^3 \gg \delta V_{pr}^3$. In this 
limit eq.~(\ref{dv_gen_k62_ext}) reduces to
the classical Kolmogorov-Obhukhov similarity,
\be
\label{rksh_tr}
    <\delta V_{tr}^p> \propto <\epsilon_r^{p/3}> <\delta V_{tr}^3>^{p/3} \, .
\ee
This equation is plotted for $p=6$ on the right of 
figure \ref{old_new}, where it is shown to hold
below the shear scale as indicated by the solid line.
Equation~(\ref{rksh_tr}) still involves integrals over the sphere. 
It may be recast into the familiar form involving only the longitudinal increments, 
evaluated in the streamwise direction $\delta u$. 
This is done by assuming that, as in the homogeneous isotropic case,
$\delta V_{tr} \simeq A(r) \delta u$ where $A(r)$ is a suitable non fluctuating
function. 
Hence eq.~(\ref{rksh_tr}) is re-written as 
\be
\label{rksh_old}
    <\delta u^p> \propto <\epsilon_r^{p/3}> <\delta u^3>^{p/3} \, .
\ee
On the contrary in the range of scales
$r>L_s$ the scale-by-scale budget shows that
the only productions terms in eq.~(\ref{dv_gen_k62_ext}) are relevant.
In this range we obtain asymptotically
\be
\label{rksh_pr}
    <\delta V_{pr}^p> \propto <\epsilon_r^{p/3}> <\delta V_{pr}^3>^{p/3}.
\ee
Equation~(\ref{rksh_pr}) is also plotted for $p=6$ on the right of figure \ref{old_new} and 
manifestly holds at the largest scales.
Equation~(\ref{rksh_pr}) exactly  corresponds to the similarity law originally
proposed by \cite{PF_mamma}.
Actually, on dimensional grounds, $\delta V_{pr}^3$ may be approximated
in terms of longitudinal and transverse velocity increments in the streamwise direction
$\delta V_{pr}^3 \simeq B(r) \delta u \delta v$ where $B(r)$ is, again, a non fluctuating 
function.  Hence, by assuming that $\delta u \delta v \simeq \delta u^2$, the similarity law 
\bfi[t!]
   \centerline{
   \epsfig{figure=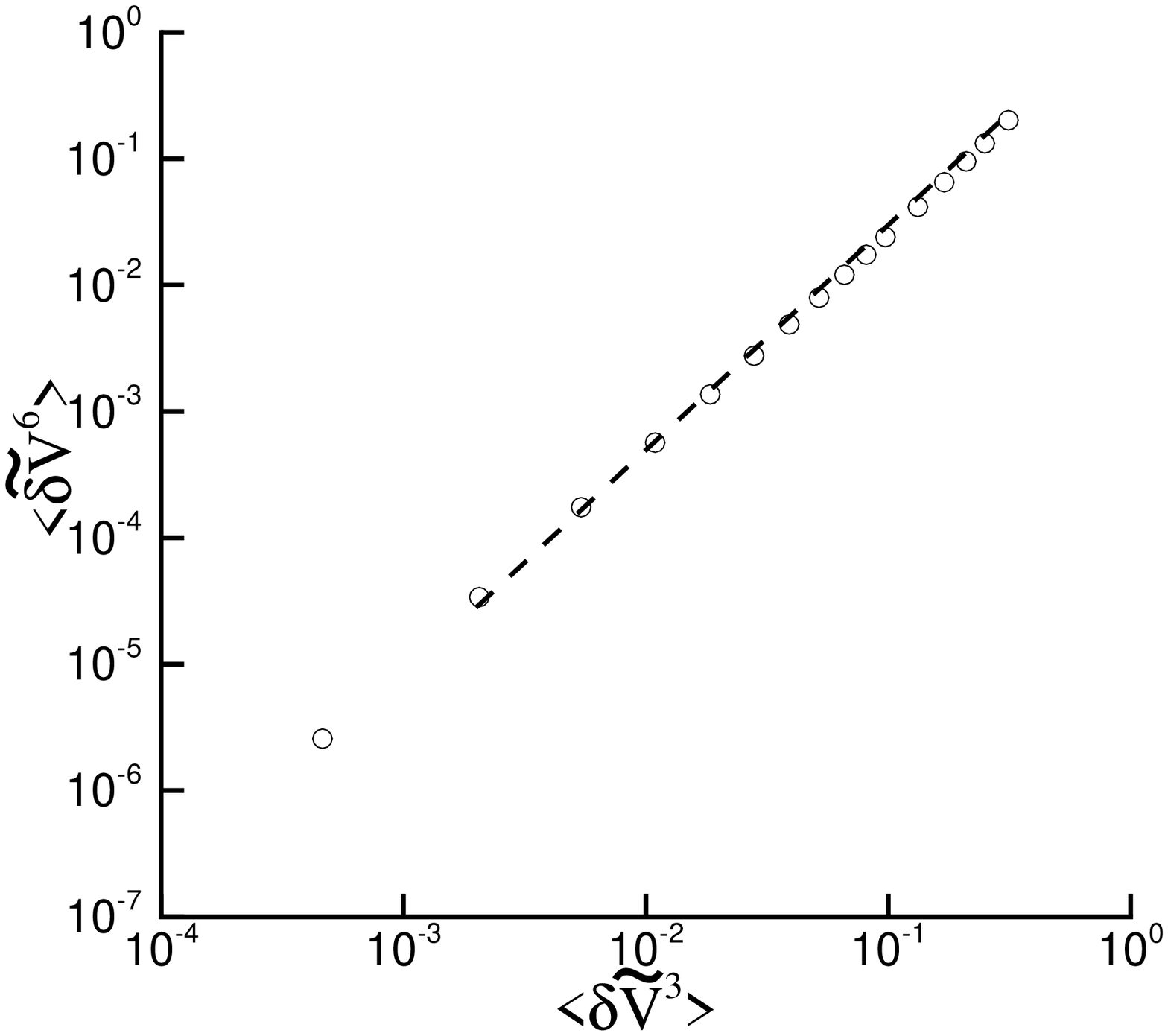,width=7.cm}
   \epsfig{figure=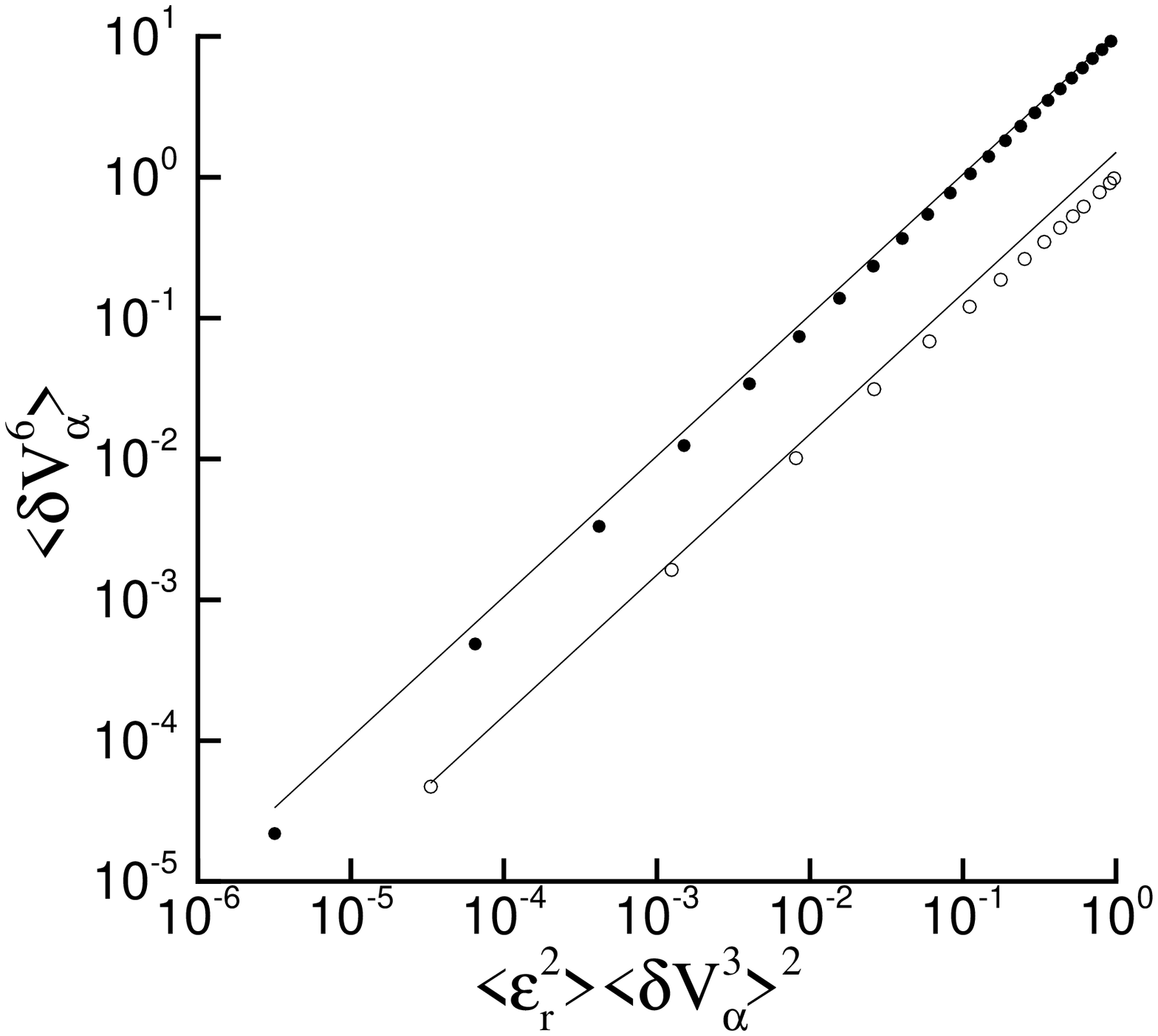,width=7.cm}
                                                      }
   \caption{Left: 
             $<\widetilde{\delta V^6}> vs. <\widetilde{\delta V^3}>$ in homogeneous isotropic  
             turbulence (dashed line with slope $s=1.78$) and in the homogeneous shear 
             flow (open circles).
             Right: 
            $<\delta V_{\alpha}^6> vs. <\epsilon_r^2> <\delta V_{\alpha}^3>^2$.
            Production ($\alpha=pr$) filled circles, transfer ($\alpha=tr$) open circles. 
            The slope of the two solid lines is $s=1$.
\label{old_new}}
\efi
\be
\label{rksh_new}
    <\delta u^p> \propto <\epsilon_r^{p/2}> <\delta u^2>^{p/2} \, 
\ee
\noi follows. As discussed in the introduction, equation~(\ref{rksh_new}) has been repeatedly
verified, both numerically and experimentally, in a number of different shear 
flows and has been shown able to correctly capture the change in the nature of the intermittency
of longitudinal increments across the shear scale.

\section{Final remarks}
The exact Karman-Howarth equation for the homogeneous shear flow has been exploited 
to provide a rational framework for the analysis of intermittency in shear turbulence, 
i.e. in flows characterized by large scale anisotropies. 
Besides confirming that production of turbulent kinetic energy prevails, as expected, at  
large scales while transfer is crucial at the smallest ones and that the two ranges are 
separated by the shear scale $L_s$, the approach has proven fruitful in allowing
the generalization of the Kolmogorov-Obhukhov theory of intermittency to shear dominated
turbulence.
In particular the two main achievements of the paper are summarized as follows.
A generalization of the structure functions can be introduced in shear turbulence, 
to extend the classical description in terms of velocity increments used in the 
standard theory for isotropic flows. 
The generalized structure functions obey a similarity law which, formally, keeps the
structure of the Kolmogorov-Obhukhov similarity, in such a way that in the limit
of zero shear the classical results of isotropic turbulence are recovered.
This generalized similarity law has been favorably checked against DNS data of a 
homogeneous shear flow and it has been shown to be consistent with both the classical 
refined similarity at small scale and the new form of similarity law recently introduced 
for shear dominated turbulence (\cite{PF_mamma}).  
While scaling laws in terms of traditional longitudinal
structure functions display two different forms of intermittency across the shear scale
(\cite{PRE}), the generalized similarity described in this paper holds uniformly across 
the entire range of available scales. 

As a further contribution, the properties of the dissipation field related to
intermittency have been shown to be  unaltered by the shear,
i.e. they are identical to those of isotropic turbulence.
This implies that the intermittency of the dissipation field displays universality,
which entails universality of the generalized velocity increments we have defined.
By exploiting the relationship between the generalized and the standard 
velocity increments we find that the universal intermittency corrections for
the generalized increments give reason for the increased intermittency of the standard velocity
increments, in agreement with previous numerical and experimental results in shear turbulence.
In particular, since the longitudinal structure functions of order $p$ correspond to 
generalized moments of the order $p$ and $3/2 \, p$ below and above the shear scale, respectively,
the flatness factor for the standard increments increases across the shear scale as a consequence
of the universality manifested by the generalized structure functions.

In conclusion we like to point the attention of the reader on the potential interest of
the present work, providing a uniform description for both shear dominated and
isotropic turbulence, in the context of LES of wall bounded flows.
Actually the present description
provides an appropriate estimate for the energy flux in non isotropic conditions, a crucial
point in developing working closure in LES ever since the now classical model 
of Smagorinski.
It could be also worth emphasizing the role of the homogeneous shear flow as the proper 
setting to check possible alternative formulations of LES closures. 


\end{document}